\documentclass[preprint2]{aastex6}

\shorttitle{ALMA-SZ shock in El Gordo}
\shortauthors{Basu et al.}
\usepackage{graphicx}
\usepackage{hyperref}
\usepackage{float}
\usepackage{psfrag}
\usepackage{amsmath}
\usepackage{amssymb}
\usepackage{ulem}
\usepackage{ctable}
\usepackage{txfonts}
\usepackage{enumerate}
\usepackage{url}
\usepackage{color}

\newcommand{\mach}{{\cal M}}

\newcommand{\plk}{{\it Planck~}}
\newcommand{\chan}{{\it Chandra~}}
\renewcommand{\deg}{{^{\circ}}}
\newcommand{\amin}{{^{\prime}}}
\newcommand{\asec}{{^{\prime\prime}}}

\defcitealias{Lind14}{L14}
\defcitealias{Men12}{M12}

\begin{document}

\title{ALMA-SZ Detection of a Galaxy Cluster Merger Shock at Half the Age of the Universe}


\author{K. Basu$^{1,\star}$, M. Sommer$^1$, J. Erler$^1$, D. Eckert$^2$, F. Vazza$^3$,  B. Magnelli$^1$, F. Bertoldi$^1$, and P. Tozzi$^4$} 

\affil{$^1$Argelander Institut f\"ur Astronomie, Universit\"at Bonn, Auf dem H\"ugel 71, 53121 Bonn, Germany; 
$^{\star}$\url{kbasu@astro.uni-bonn.de} \\
$^2$ Department of Astronomy, University of Geneva, Chemin d'Ecogia 16, 1290 Versoix, Switzerland \\
$^3$ Hamburger Sternwarte, Gojenbergsweg 112, 21029 Hamburg, Germany \\
$^4$ INAF$-$Osservatorio Astrofisico di Arcetri, Largo E. Fermi 5, I-50125 Firenze, Italy}

\begin{abstract}

We present ALMA measurements of a merger shock using the thermal Sunyaev-Zel'dovich (SZ) effect signal, at the location of a radio relic in the famous El Gordo galaxy cluster at $z \approx 0.9$. Multi-wavelength analysis in combination with the archival \chan data and a high-resolution radio image provides a consistent picture of the thermal and non-thermal signal variation across the shock front and helps to put robust constraints on the shock Mach number as well as the relic magnetic field. We employ a Bayesian analysis technique for modeling the SZ and X-ray data self-consistently, illustrating respective parameter degeneracies.  Combined results indicate a shock with Mach number $\mach = 2.4^{+1.3}_{-0.6}$, which in turn suggests a high value of the magnetic field (of the order of $4-10 ~\mu$G) to account for the observed relic width at 2 GHz.  At roughly half the current age of the universe, this is the highest-redshift direct detection of a cluster shock to date, and one of  the first instances of ALMA-SZ observation in a galaxy cluster. It shows the tremendous potential for future ALMA-SZ observations to detect merger shocks and other cluster substructures out to the highest redshifts. 

\end{abstract}


\section{Introduction} 
\label{sec:intro}

Intergalactic shocks created by galaxy cluster mergers 
are among the most spectacular events in the structure formation history of the universe.  
They raise the thermal energy of the intracluster medium to the keV range and also accelerate a population of seed electrons to relativistic energies (e.g., \citealt{Sar02}, \citealt{Brug12}). 
These GeV energy electrons are believed to be responsible for producing the megaparsec-scale diffuse synchrotron sources  known as radio relics (\citealt{Enss98}, \citealt{Nuz12}, \citealt{Skill13}).
While the connection between merger shocks and the non-thermal relic emission has been established in the low-redshift universe 
through X-ray observations (e.g., \citealt{Fin10}, \citealt{Aka13}), at high redshifts it becomes extremely difficult due to the dimming of the X-ray signal. In this regard, the Sunyaev-Zel'dovich (SZ) effect (\citealt{SZ72}) can be an ideal alternative since its brightness is redshift independent, opening a new window for observing cluster shocks across the visible universe. 
The first use of the SZ effect to model a relic shock was done recently for the Coma cluster using \plk data (\citealt{Er15}). In this Letter, we present the first SZ result from ALMA for a merger shock in the famous El Gordo cluster. 

The galaxy cluster ACT-CL J0102$-$4915 at $z=0.87$, nicknamed ``El Gordo,'' is the most massive high-redshift cluster known and was the most significant SZ detection in the ACT survey from where it was discovered (\citealt{Men10}, \citealt{Marri11}). 
\citealt{Men12} (hereafter \citetalias{Men12}) described the merging nature of this cluster in a multi-wavelength analysis, and  
\citealt{Lind14} (hereafter \citetalias{Lind14}) presented its non-thermal environment from GMRT and ATCA data, 
discovering a giant radio halo and a set of peripheral radio relics. These are the highest redshift radio halo and relics known and are among the most luminous. Our ALMA observation targeted the prominent NW relic with a 3 hr exposure at 100 GHz, 
detecting the sign of an underlying pressure discontinuity.

\begin{figure*}
\hspace*{-2mm}
  \includegraphics[width=9.4cm, height=9.4cm, clip=true, trim=10 10 60 10]{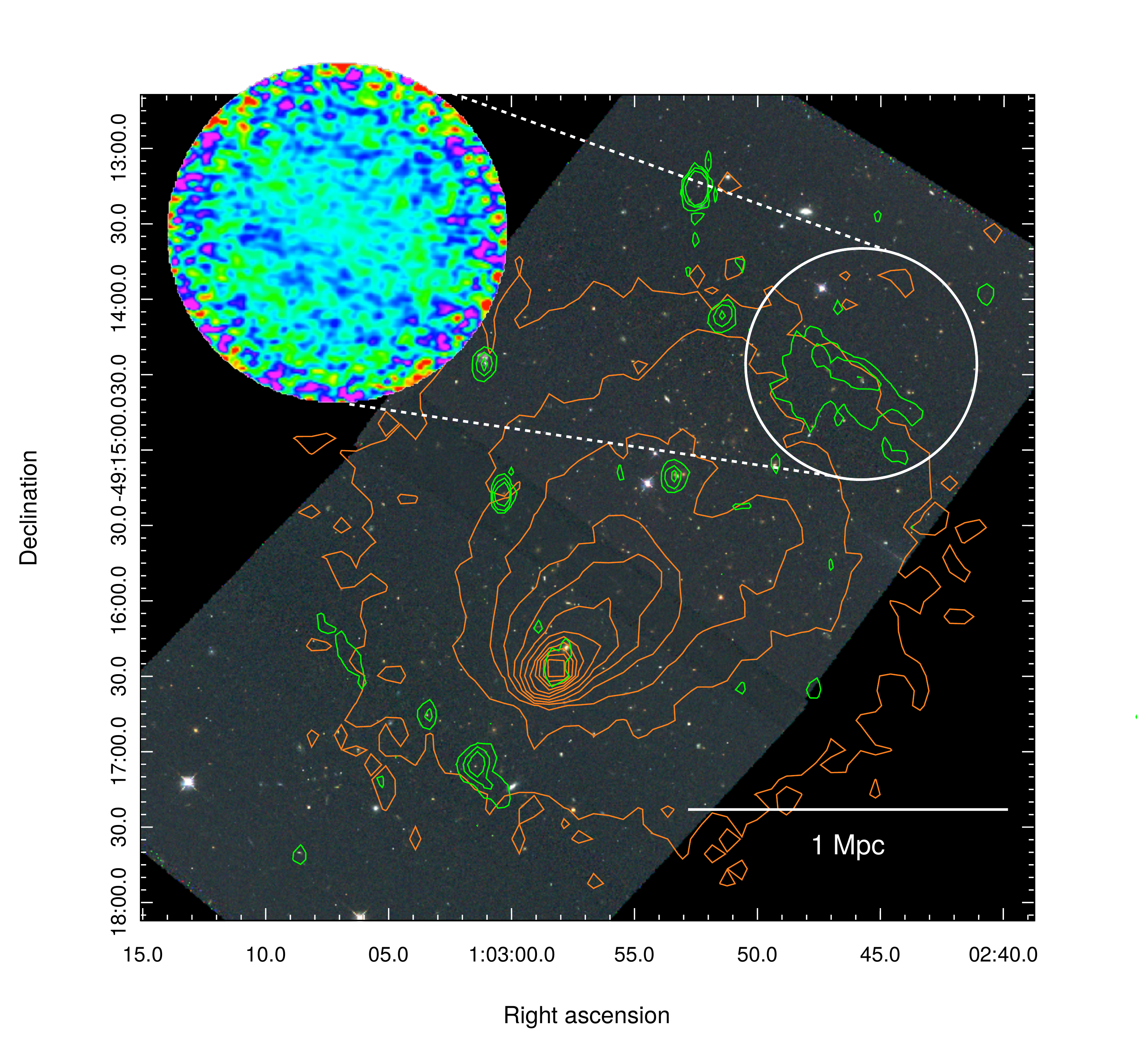}%
  \hspace{-1mm}
  \includegraphics[width=10.1cm, height=8.8cm, clip=true, trim=40 20 30 50]{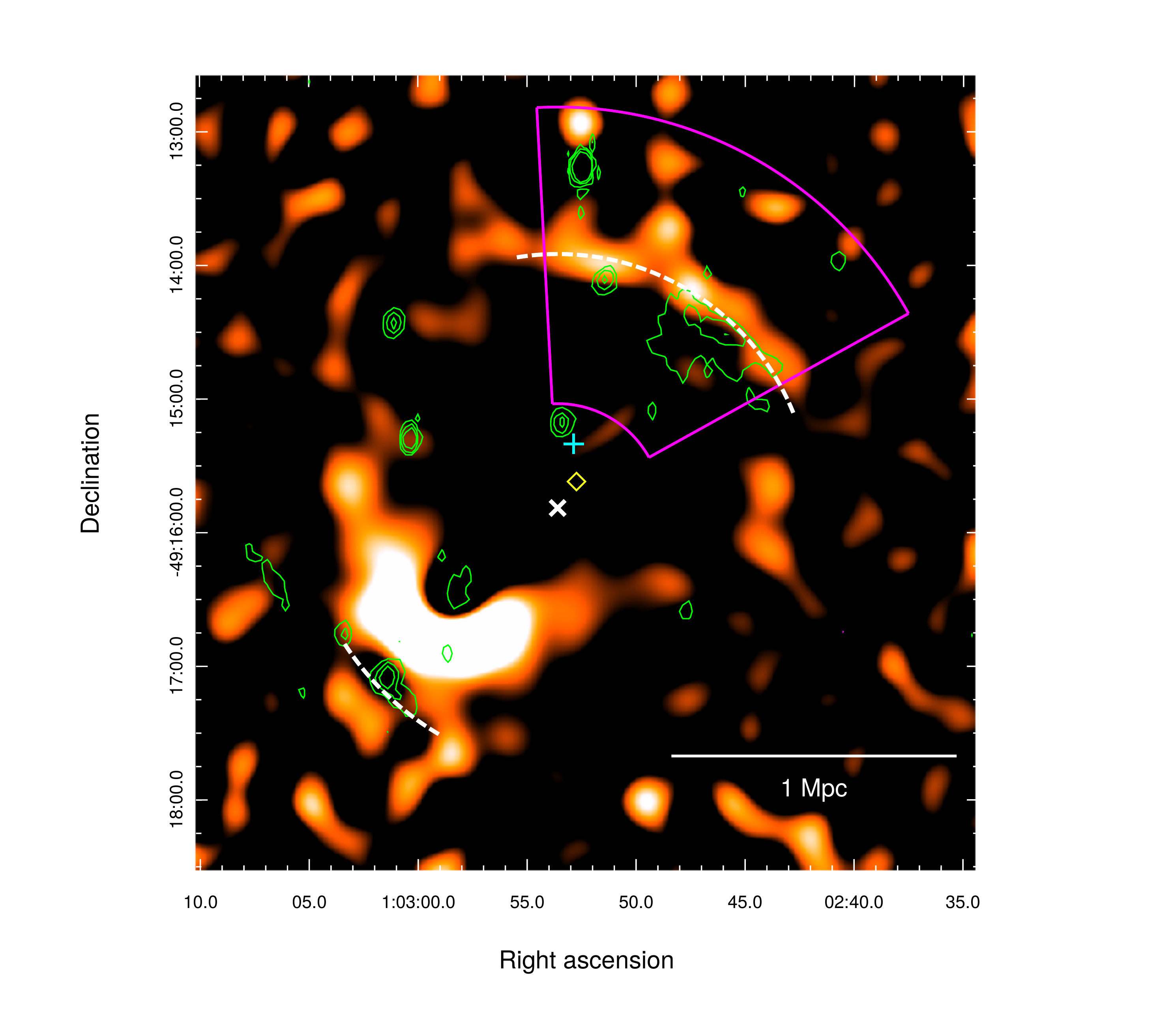}  
  \caption{Multi-wavelength view of the El Gordo cluster and its NW relic. {Left:} a color-composite image made from two HST/ACS pointings, overlayed with contours from the soft-band \chan (orange)  
  and the 2.1 GHz ATCA radio continuum (green) data. The white circle marks the region imaged by ALMA, with a zoomed-out inset showing the flux distribution as obtained from imaging the ALMA data (see Fig. \ref{fig:alma} for scales). 
  {Right:} an unsharp-masked image of the X-ray brightness in color, highlighting an arc-like shock front 
  extended well beyond the radio synchrotron emission (green contours).  The magenta sector marks the area used for the main X-ray analysis.  
  White-dashed arcs mark the tentative shock fronts, whose center is the white `$\times$' in the middle defining the center of our shock model. The blue `$+$' is the ACT-SZ centroid (\citealt{Marri11}), and the yellow `$\Diamond$' is the weak-lensing center of mass (\citealt{Jee14}).}
  \label{fig:all}
\end{figure*}

The presence of a shock at El Gordo's NW relic location is already hinted at by the observation of a radio spectral index gradient, 
from which \citetalias{Lind14} derived a shock Mach number $\mach = 2.5^{+0.7}_{-0.3}$. 
Recently, \citet{Bott16} reported on modeling the same 360 ks \chan archival data that we use in this Letter, inferring a strong shock ($\mach \gtrsim 3$) from a brightness discontinuity measurement. 
However, X-ray brightness edges can also be the result of cold fronts that can extend to the cluster outskirts in large-scale gas sloshing \citep[e.g.,][]{Ross13}. Measuring a pressure discontinuity through the SZ effect, on the other hand, provides an unequivocal case for a shock. 
Accurate shock Mach number determination also helps to infer the infall velocities of the merging subclusters, and such velocity tests are a critical tool in examining current cosmological models, particularly with high-$z$ systems like El Gordo (e.g., \citealt{Lee10}, \citealt{Katz13}, \citealt{Mol15}). 
This makes ALMA, with its superb angular resolution and sensitivity, as one of the forefront instruments for probing cluster astrophysics and cosmology through SZ shock modeling.

This Letter presents a concise description of the multi-wavelength data and their analysis methods (Section \ref{sec:data}), results for this highest-redshift shock (Section \ref{sec:res}), and a summary (Section \ref{sec:dis}). More details on the SZ and X-ray analysis, supplemented by upcoming ALMA Compact Array data, will be presented in a future paper. 
We use $\Lambda$CDM cosmology parameters $\Omega_{m}=0.27$, $\Omega_{\Lambda}=0.73$ and $H_0=70$ km s$^{-1}$ Mpc$^{-1}$ (\citealt{Hin13}); thus, $1\amin$ corresponds to 471 kpc at El Gordo's redshift. 
Parameter values are quoted with 68\% credible intervals, unless otherwise noted.


\section{Data and Analysis Methods}
\label{sec:data}

Fig. \ref{fig:all} gives an overview of the X-ray and radio data available for the El Gordo cluster and puts our ALMA observation into perspective. The left panel shows an RGB-color image made from two {\it Hubble Space Telescope} (HST) pointings spanning the NW$-$SE direction, overlayed with the $0.5-2$ keV X-ray (orange) and 2.1 GHz radio (green) contours. The field of view of ALMA at 100 GHz is marked by the white circle (primary beam FWHM $\sim 1\amin$). In the right panel, an unsharp-masked (``edge-detected'') X-ray image, created from differencing two images smoothed by $\sigma=25\asec$ and $\sigma=8.5\asec$ resolutions, describes the shock geometry. 
 There is a large ($\sim$ 1 Mpc length) arc-like shock feature in the NW direction that extends well beyond the relic region (see also \citetalias{Men12}). Two opposing radio relics in the NW and SE directions suggest a merger in the plane of the sky along their common axis (\citetalias{Lind14}, \citealt{Ng15}). We take the midpoint of these arcs, at R.A., decl. $=[15.723\deg,-49.264\deg]$, as the center of our spherical shock model (marked by a white `$\times$'). 
 The shock radius, here the same as its radius of curvature, is fitted individually from each data set.

\begin{figure*}[t]
\hspace*{-5mm}
  \includegraphics[height=4.9cm]{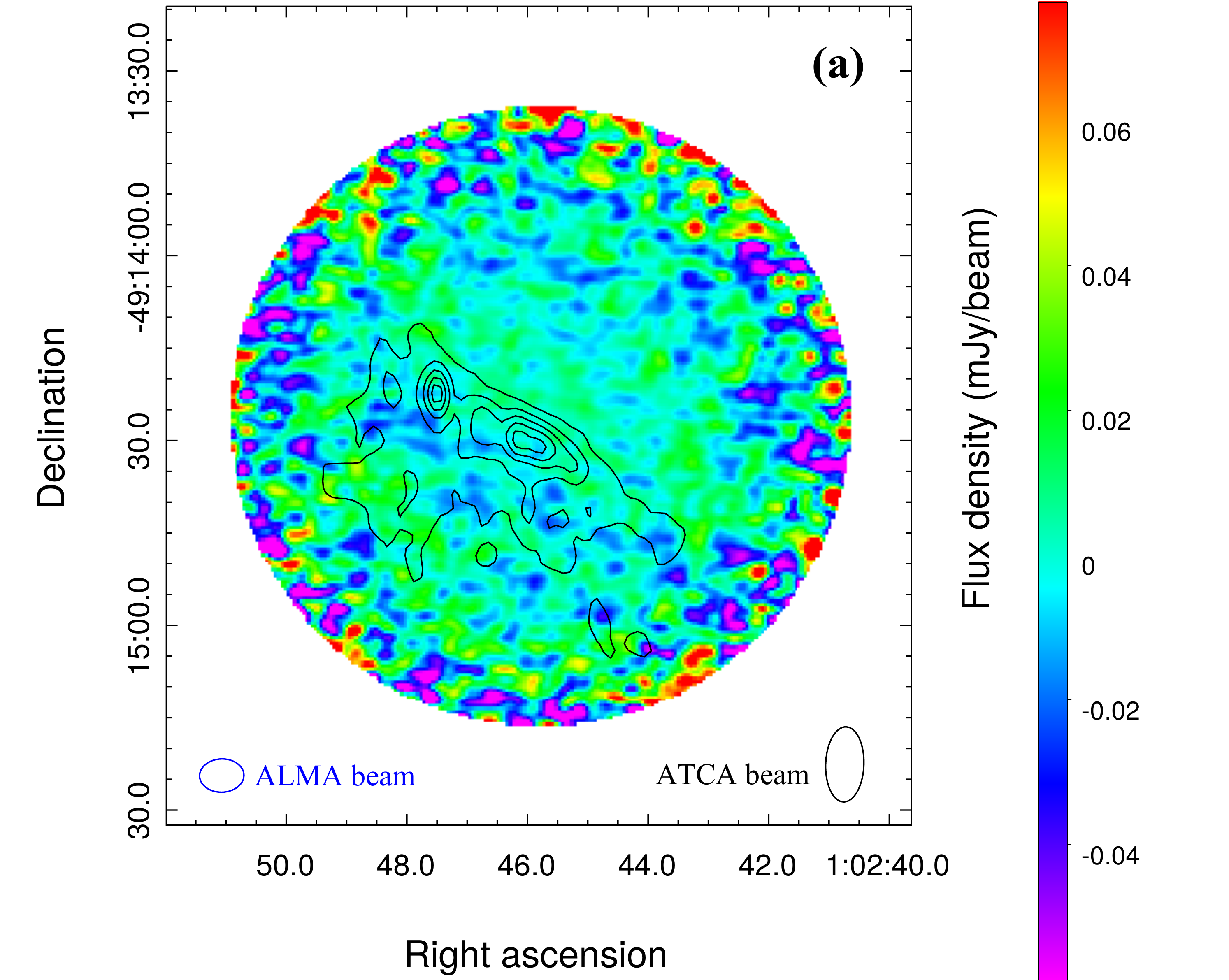}%
  \hspace*{-4mm}
  \includegraphics[height=4.9cm]{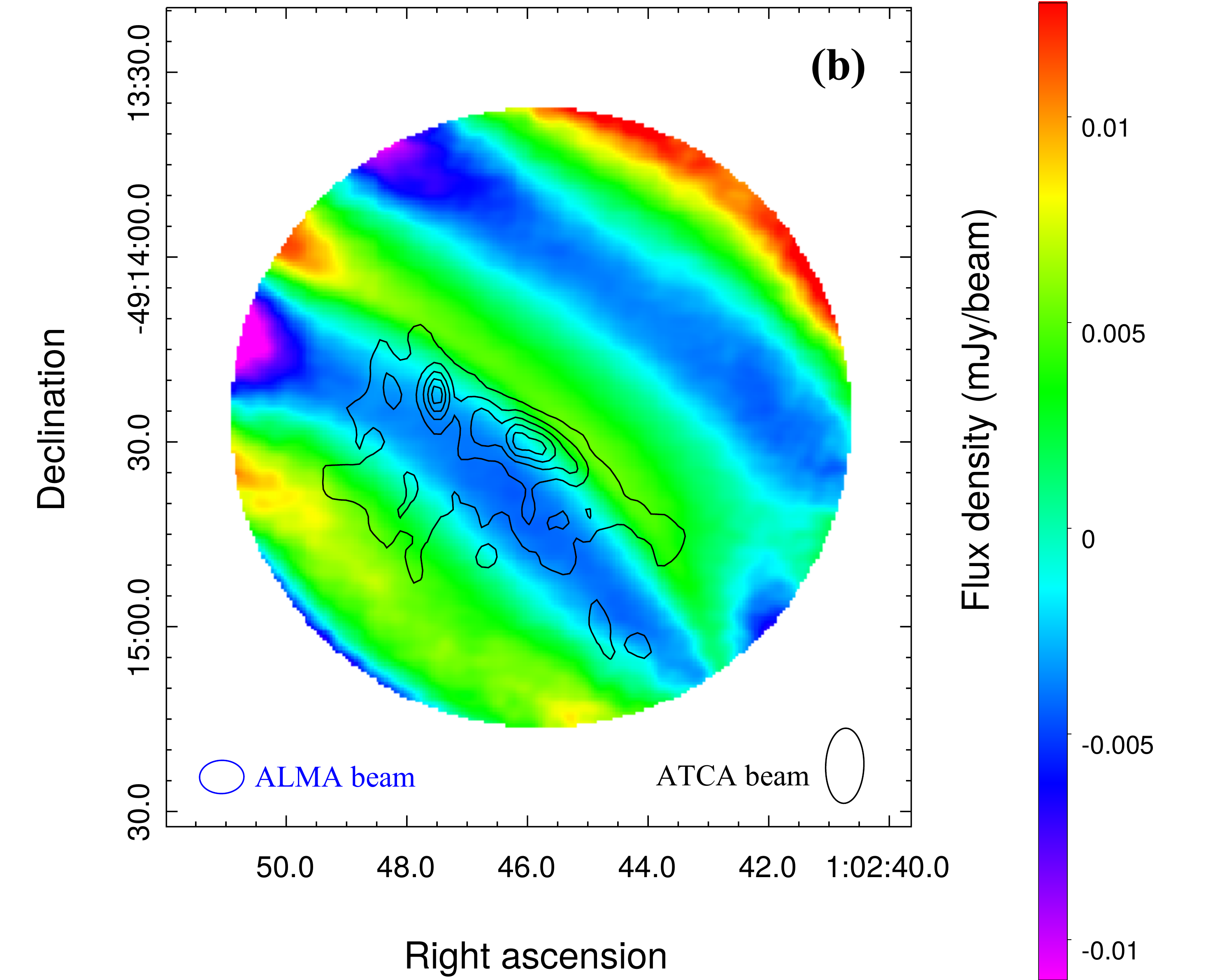}%
  \hspace*{-5mm}
  \includegraphics[height=5.6cm]{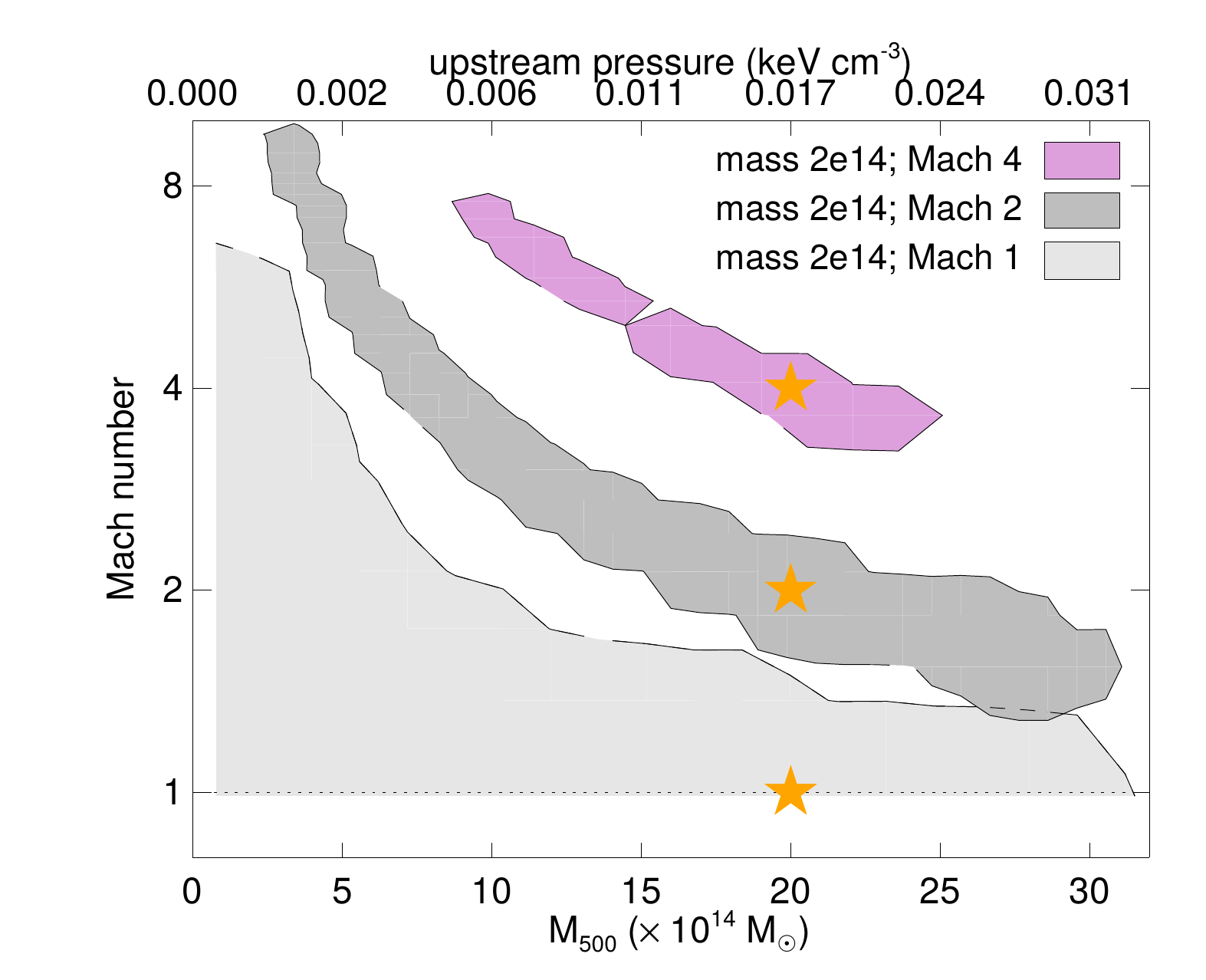}
  \caption{Real and simulated ALMA observations. {From left:} direct deconvolved or `dirty' image made from the real ALMA data (a), and a simulation with negligible noise for a $\mach=2$ shock to highlight the ripple-like signal (b).  
  The radio relic contours at 2.1 GHz are shown for reference.  {Right:} results for model fitting from mock ALMA observations with realistic noise, for a ${\cal M} =$ 1 (no shock), 2 or 4 shock in a massive El Gordo like cluster.  
  Contours define the 95\% posterior regions,  and the input values are indicated by the stars. 
  ALMA interferometric observation is insensitive to any large-scale SZ signal and detects primarily the pressure difference 
 ($\Delta p$) across the shock.}
  \label{fig:alma}
\end{figure*}

\vspace{-2mm}
\subsection{ATCA 2.1 GHz radio data}

The radio data analysis is described in \citetalias{Lind14}, we use a high-resolution total intensity image  made from the ATCA 2.1 GHz observation.  
The effective resolution is $6.1\asec \times 3.1\asec$, making the relic sufficiently resolved in the direction of shock propagation to fit a radial brightness profile. 
We use a lognormal emissivity model (\citealt{szrelic}) that is projected from a 3D  cone and smoothed to the ATCA resolution. 
The fit is shown in Fig. \ref{fig:prof} (top panel), where the error-bar (lower-right) represents the noise at the phase center.  

\subsection{\chan X-Ray data}

We use archival \chan data for El Gordo totaling 360 ks of observation (ObsID: 12258, 14022, 14023). The data are reduced with the  
\textsc{CIAO} software (\textsc{CALDB} 4.7.1)  
to produce calibrated image and background files from each ObsID individually and then are added together weighted by the respective exposures. 
We define a ``shock-cone'' that encompasses the entire arc-like shock feature, 
between $26\deg$ and $94\deg$ angles (magenta sector in Fig. \ref{fig:all}, right), 
to maximize the cluster signal-to-noise (S/N) ratio in the outer region. We 
also model the data separately in the radio relic sector (between $26\deg$ and $64\deg$ angles) and its complement. 

The X-ray brightness is modeled with a standard broken power law in density \citep[e.g.,][Eqn. 1]{Eck16} 
using six parameters: the shock radius, pre-shock electron density, upstream and downstream density slopes, the Mach number, and the background level (marginalized over 5\% uncertainty).  
The shock density jump is derived from the standard Rankine-Hugoniot  (R-H) condition (with adiabatic index 5/3), as $n_{\mathrm{down}}/n_{\mathrm{up}} = 4\mach^2 / (\mach^2+3)$, which is valid for $\mach \geq 1$. 
This model is projected assuming a spherical geometry and fit to the $0.5-2$ keV brightness profile using a Markov Chain Monte Carlo (MCMC) method. El Gordo is a complex system, and the gas morphology in the NW sector is particularly disturbed, as \citetalias{Men12} has pointed out a low-density 
``wake" in this direction. To keep our spherical projection approximately valid, we limit the fit to 400 kpc downstream.  
Spectral fitting was done in the $0.5-7$ keV range using a single-temperature model.

\subsection{ALMA-SZ data}

The relic was observed by the ALMA main array in Cycle 3 during 2015 December, employing thirty-five 12-meter antennas in the most compact configuration (2015.1.01187.S; PI: K. Basu). Four continuum bands were centered at 93, 95, 105 and 107, GHz, covering a total 7.5 GHz bandwidth. Total duration of the observation was 5.2 hr, which after calibration yielded roughly 3 hr of on-source data. We used the calibrated data products obtained with the \textsc{CASA} software as provided by the ALMA project, with a few additional flaggings. 
Imaging using CLEAN with natural weighting resulted in a noise rms of $6~ \mu$Jy/beam at the phase center, for a synthesized beam size of $3.6\asec \times 2.7\asec$ (P.A. $88\deg$). 
There is no evidence for point-like sources in the ALMA image.

A CLEAN image example is shown in the Fig. \ref{fig:all} inset, and a direct deconvolved image (``dirty-image'') in Fig. \ref{fig:alma} (left panel); the difference is negligible due to the low S/N. These images are shown only for illustration; we \textsl{do not} use any imaging product for the shock modeling, rather fit our model directly to the ALMA $uv$-data after Fourier transforming and de-gridding it into the visibility plane. This is done using an MCMC method within \textsc{CASA} and is computationally expensive, due to the large data volume. 
The $93-107$ GHz SZ spectrum is computed with relativistic corrections (\citealt{Ito04}) due to the high post-shock temperatures. 

The $uv$-fitting method safeguards against possible biases that can occur when imaging a diffuse, low S/N negative signal like a cluster shock in SZ. We make extensive simulations for mock ALMA observations, and 
 some results are shown in Fig. \ref{fig:alma}. The middle panel is a dirty image from a practically noiseless simulation, highlighting the ripple-like intensity pattern which is a deconvolution artifact (see also Fig. \ref{fig:prof}, bottom panel). 
Fig. \ref{fig:alma} (right) shows fit results for realistic noise, 
simulating a $\mach =1$, 2, or 4 shock model on a massive El Gordo-like cluster. 
The noise is obtained from randomizing the phases in individual scans. 
In the $\mach=1$ case (no shock), ALMA filters out the entire large-scale SZ signal from the cluster, and the measurement is consistent with noise. For a weak shock ($\mach=2$) ALMA clearly detects a pressure discontinuity, but its amplitude is anti-correlated with the pressure normalization. This is an inherent limitation of the current ALMA observation (`short-spacing' problem).  
With higher $\mach$-values the S/N improves, 
lowering the uncertainties but not removing this degeneracy. 

To fit the shock, we choose a spherical GNFW pressure model for the cluster (\citealt{Nag07}, \citealt{Ar10}),  
adding a pressure boost at the shock radius from the R-H condition, as $p_{\mathrm{down}}/p_{\mathrm{up (GNFW)}} = (5\mach^2 -1)/4$. The downstream and upstream slopes are taken to be consistent with the X-ray results, 
leaving the pressure amplitude, shock radius and the Mach number as free parameters.  
Since X-ray data indicate a shallow upstream density profile (Sec. \ref{sec:xmodel}), we choose GNFW model parameters with a shallow outer pressure slope \citep[][]{Say13} and marginalize its value within 30\% uncertainty.  
The small field of view of ALMA makes our modeling insensitive to any parameter choice that regulates 
the pressure profile shape far away from the shock.


\section{Results and Discussion}
\label{sec:res}

\begin{figure}
\hspace*{-2mm}
  \includegraphics[width=1.0\columnwidth]{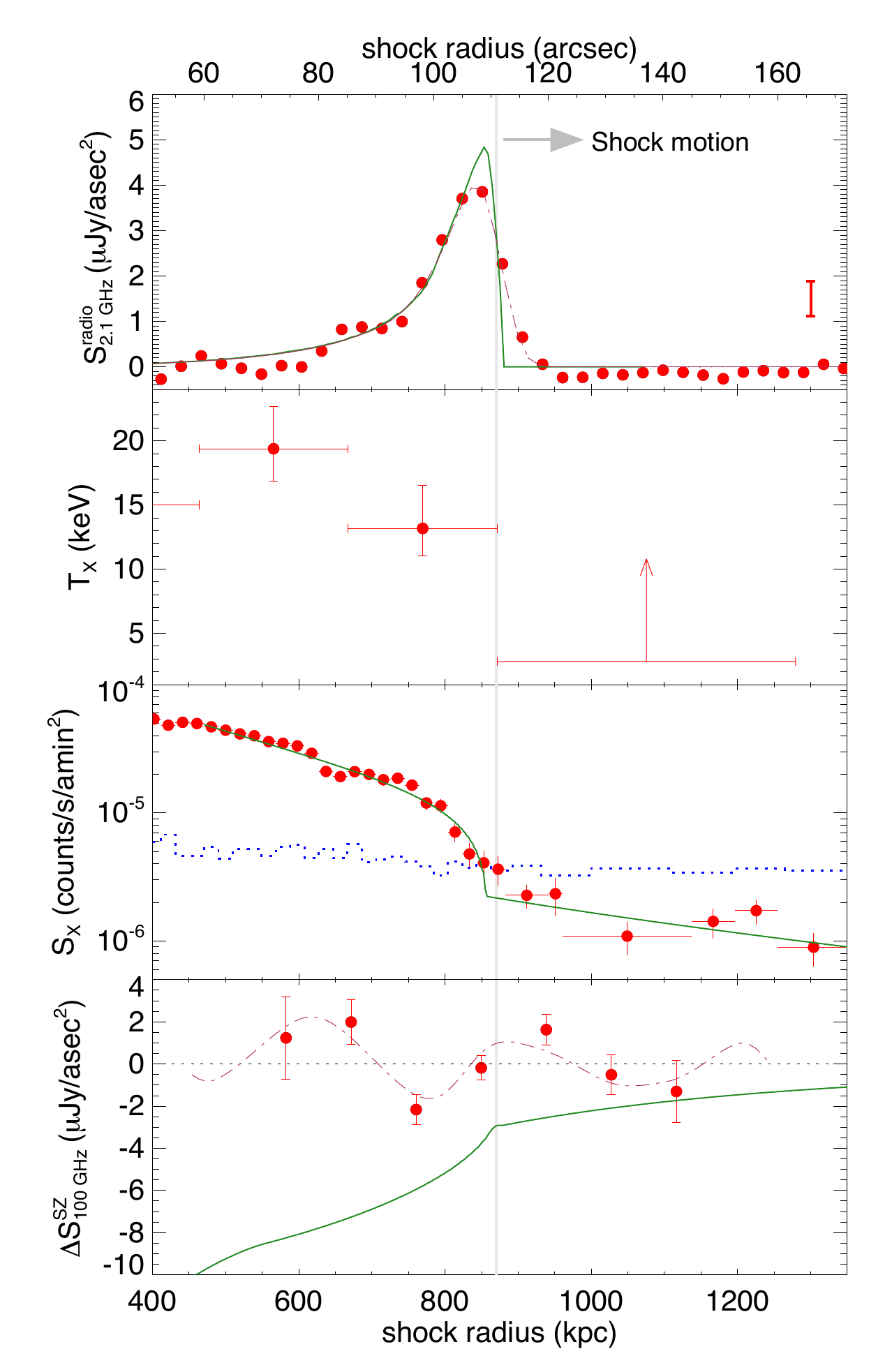}
  \caption{Thermal and non-thermal signal variations across the relic shock. {From top:} radio synchrotron emission at 2.1 GHz, X-ray temperature measurements from the {\it Chandra} $0.5-7$ keV data, X-ray surface brightness in the $0.5-2$ keV band, and the SZ flux modulation at 100 GHz as observed by ALMA. The green lines show the respective best-fit theoretical models, and the red dotted-dashed lines are the observed profiles after beam smoothing (radio) or image deconvolution (SZ). The X-ray temperature lower limit is at the 90\% confidence level, and the blue dotted line in the third panel marks the mean X-ray background. The vertical gray line is the best-fit shock location derived from  the SZ data.}
  \label{fig:prof}
\end{figure}

Fig. \ref{fig:prof} is a summary of our multi-wavelength analysis, in a first modeling of  the radio synchrotron, X-ray temperature and brightness, and  the thermal SZ effect signal across a cluster shock. Even with a deep ($\sim 100$ hr) \chan observation, the pre-shock temperature in this high-$z$ object remains unconstrained due to the high background-to-cluster photon ratio. The X-ray background level is shown by the blue dotted line in the third panel.  
In the bottom panel, the \textsl{observed} SZ signal is the ripple-like feature (red dotted-dashed line), also seen in Fig. \ref{fig:alma}.

\begin{figure*}
\hspace*{-5mm}
  \includegraphics[width=1.1\columnwidth, height=6.4cm]{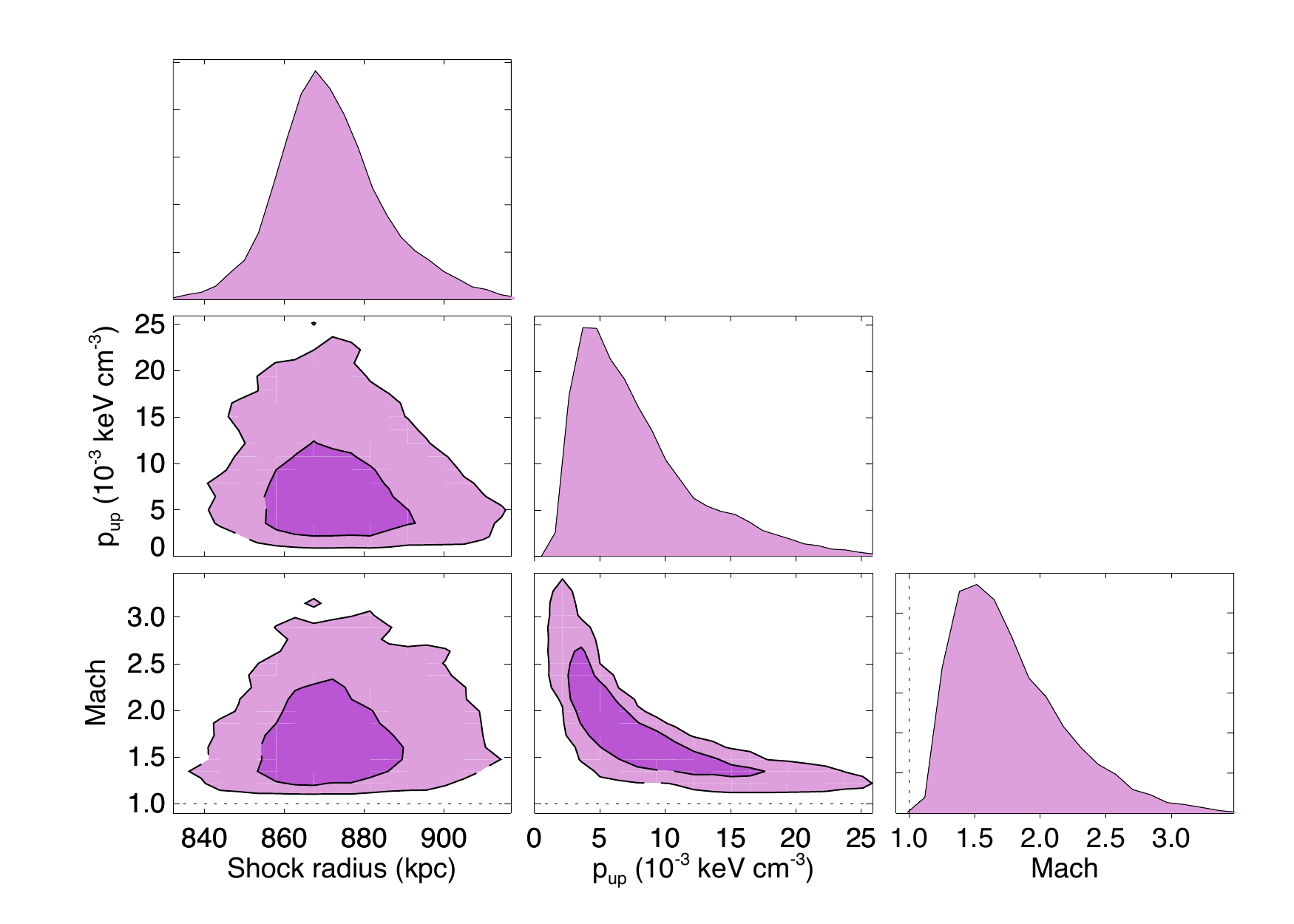}%
  \hspace*{-4mm}
  \includegraphics[width=1.1\columnwidth, height=6.4cm]{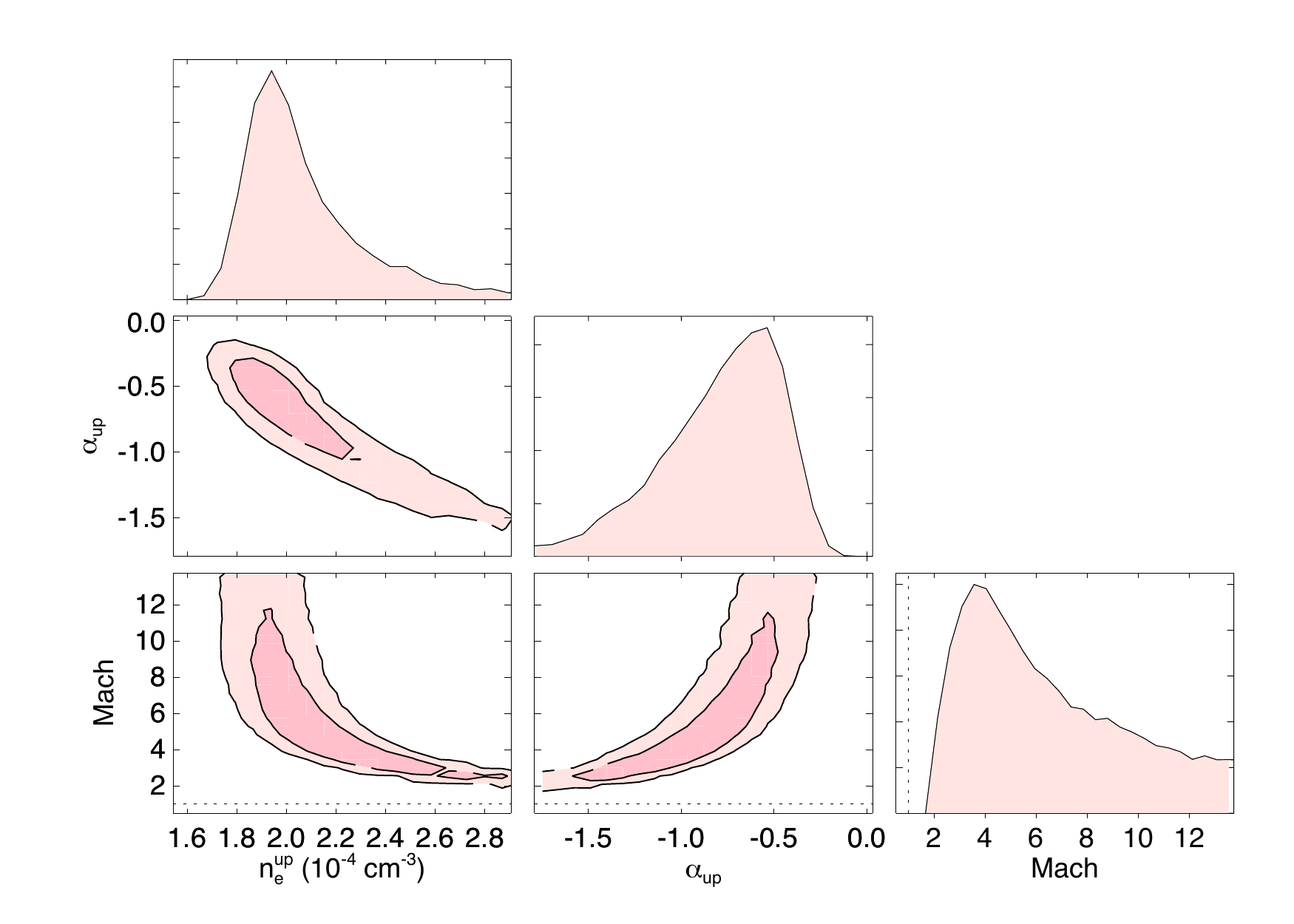}
  \caption{SZ and X-ray modeling results, showing the 68\% and 95\% posterior probabilities (darker and lighter contours) and the marginalized parameter values.  
  {Left:} results obtained from the ALMA-SZ data, with a prior on the GNFW pressure normalization based on El Gordo's mass (Sec. \ref{sec:szmodel}). The shock location is well-constrained,  but the shock Mach number is highly anti-correlated with the upstream pressure amplitude ($p_{\mathrm{up}}$). 
  {Right:} results from the \chan X-ray brightness modeling, using an SZ prior on the shock location and a 5\% error on the background. 
  Both the pre-shock gas density ($n_e^{\mathrm{up}}$) and its power-law slope ($\alpha_{\mathrm{up}}$) have strong non-zero correlation with the Mach number.}
  \label{fig:corner}
\end{figure*}

\subsection{SZ shock modeling}
\label{sec:szmodel}

An SZ fit result example, based on the GNFW$+$shock pressure model, is shown in the Fig. \ref{fig:corner} (left panel). 
The high angular resolution of ALMA constrains the shock radius to good accuracy at $r_{\mathrm{sh}}=870^{+14}_{-13}$ kpc. This value is uncorrelated with other model parameters, and the uncertainty roughly corresponds to the effective size of the ALMA synthesized beam, approximately $3.5\asec$ (27.5 kpc) in FWHM. There is a strong anti-correlation between the Mach number and the upstream pressure amplitude, which results from the unconstrained normalization.
SZ data alone prefer a weak shock,  with a large uncertainty range for the Mach number: 
$\mach = 1.4^{+1.2}_{-0.2}$, supporting $\mach > 1$ at more than $98\%$ confidence. 
Putting a weak-lensing prior on the total mass ($M_{500,c}=1.80\pm 0.34 \times 10^{15}$ M$_{\odot}$; \citealt{Jee14}) partially breaks this degeneracy and reduces the errors to $\mach = 1.5^{+0.5}_{-0.2}$ (shown in the figure). However, for a merging cluster like El Gordo a global mass prior will be inaccurate for modeling the pressure in a disturbed sector, so  
we derive a pressure prior from the X-ray data.

\subsection{X-Ray shock modeling}
\label{sec:xmodel}

X-ray brightness modeling supports a strong shock ($\mach \sim 4$) propagating outward in a low-density region where the pre-shock medium has a  shallow density slope ($\sim r^{-0.6}$). 
Parameters are shown in Fig. \ref{fig:corner} (right panel); the Mach number is strongly degenerate with the amplitude and the slope of the pre-shock density due to the noisy upstream data. 
Moreover, the insensitivity of a density ratio measurement for $\mach \gtrsim 4$ shocks adds to the degeneracy. 
X-ray data alone prefer a smaller shock radius, at $854^{+11}_{-13}$ kpc, consistent with the SZ result but in mild tension with the radio (Sec. \ref{sec:radmodel}).  
Marginalizing over a 5\% background calibration error and placing an SZ prior on the shock radius, we obtain a Mach number $3.5^{+6.4}_{-1.3}$, downstream slope $-1.2^{+0.1}_{-0.1}$, upstream  slope $-0.6^{+0.2}_{-0.4}$, and upstream density at the shock radius $1.9^{+0.3}_{-0.2} \times 10^{-4}$ cm$^{-3}$. 
Fixing the background (i.e. no systematics) 
prefers a shallower upstream slope and hence biases the Mach number to be high. Possible inverse Compton (IC) emission in the post-shock region at this redshift can also cause a high-Mach bias. Both  downstream and upstream brightness data show hints of substructures (Fig. \ref{fig:prof}), giving a poor fit with the broken power law model and suggesting a possible clumpy medium.

Preceding X-ray results are obtained within the full $\sim70\deg$ shock-cone, and results from the relic-cone and its complementary sector are consistent.  Both show evidence for a shock, with  $\mach = 2.9^{+7.8}_{-0.9}$ and $\mach = 2.3^{+3.0}_{-0.8}$, respectively. This is the first evidence of a cluster merger shock clearly extending beyond its synchrotron emitting region. 
A possible explanation for this asymmetry could be the presence of an AGN-like source in the relic, unresolved in the ATCA image (significance below $3\sigma$), that is seeding only part of the shock with relativistic electrons and magnetic fields.

\begin{figure}
\hspace*{-3mm}
  \includegraphics[width=\columnwidth,height=6.5cm]{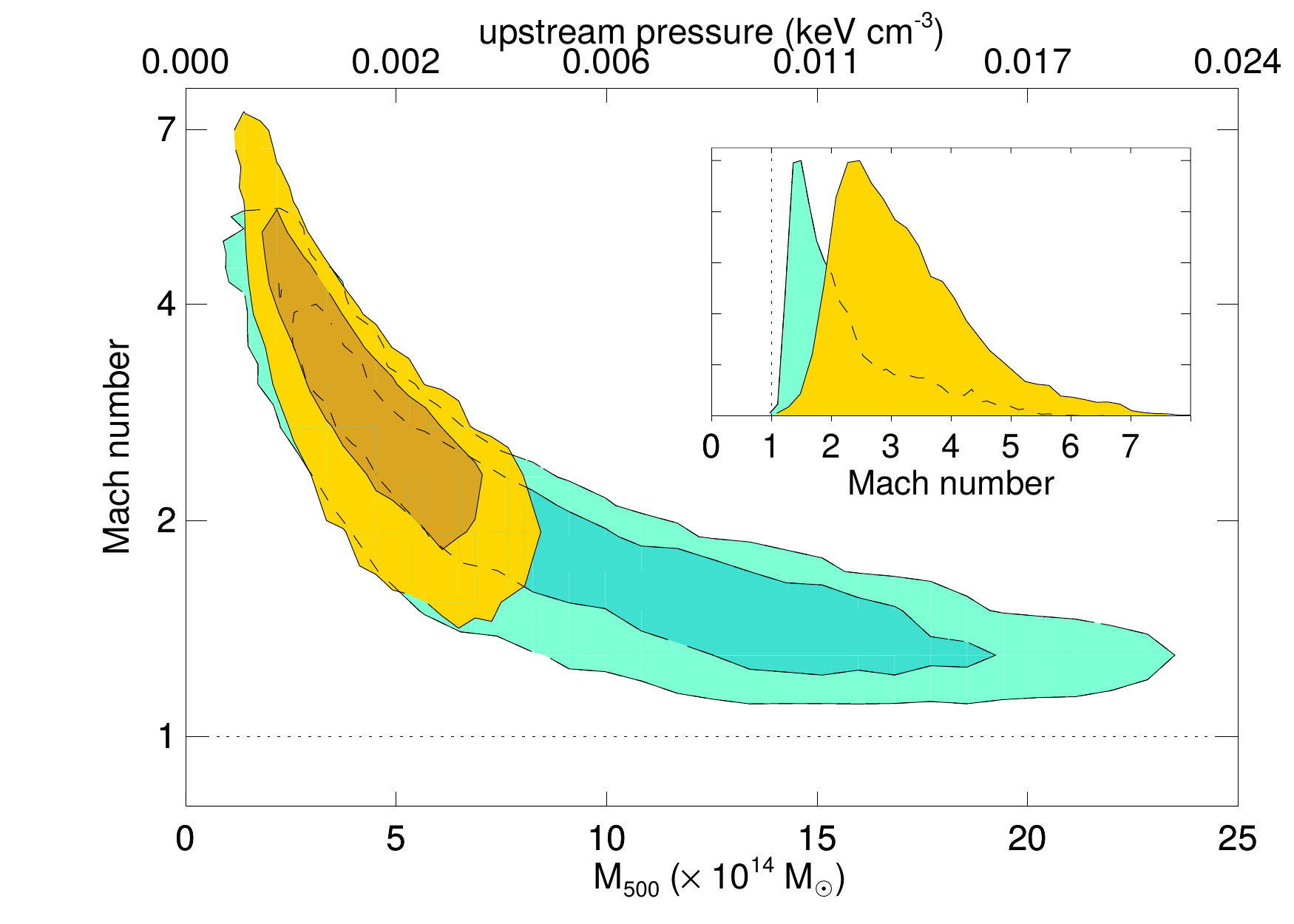}
  \includegraphics[width=\columnwidth,height=6.5cm]{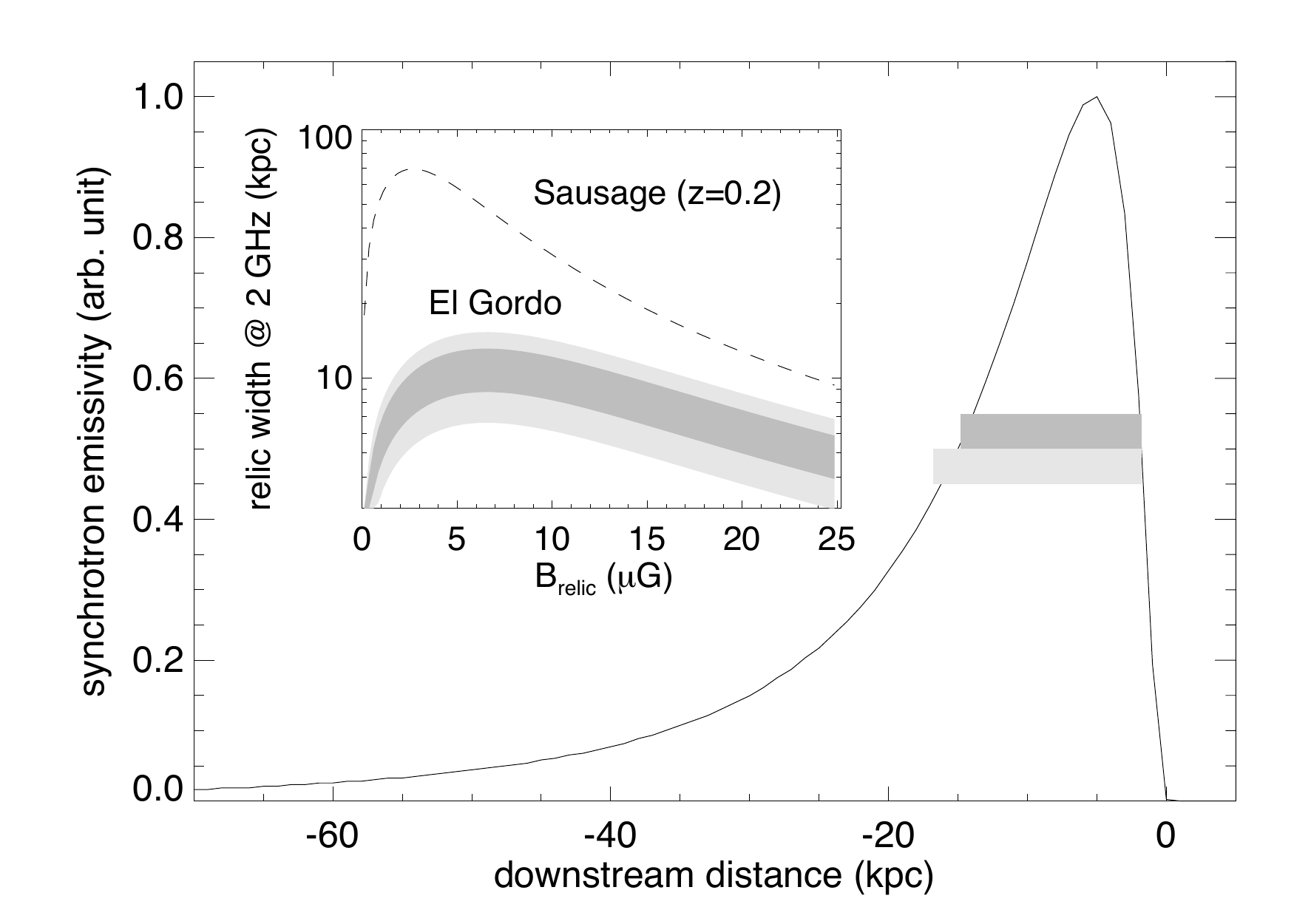}
  \caption{Constraints on the shock Mach number and the relic magnetic field. {Top:} posterior probabilities for the Mach number and cluster mass/pre-shock pressure: the green contours are from ALMA-SZ modeling only, and the yellow ones are after using an X-ray pressure prior. Darker and lighter colors mark the 68\% and 95\% credible regions. The inset figure shows the marginalized Mach number distributions. 
  {Bottom:} estimation of the relic magnetic field based on its 2.1 GHz width. The solid line shows the deprojected synchrotron emissivity profile, and the two horizontal bands mark 
  the \textsl{maximum} relic width at 68\% (dark gray) and 95\% (light gray) confidence as expected from the 
  radiative lifetime of the accelerated electrons. These limits come from the inset figure, where we show the relic width as a function of the magnetic field strength, using gas temperature and Mach number ranges derived from the joint SZ/X-ray modeling. 
  Same curve for the nearby Sausage relic is shown for comparison \citep[][]{vanWe10}. }
  \label{fig:final}
\end{figure}

\subsection{Joint SZ/X-Ray modeling}

To better constrain the SZ shock Mach number, we use a pressure prior  
assuming the upstream X-ray density and temperature measurements are independent. Since the upstream $T_\mathrm{X}$ has no effective upper bound, 
we scale from the downstream $T_\mathrm{X}$ value using the Mach number in a chain, from the R-H condition for temperature jump. The post-shock temperature is measured from a wider $1\amin$ region ($T_\mathrm{X} = 16.8^{+2.4}_{-1.8}$ keV) to avoid possible biases immediately behind the shock from an IC component, electron/ion non-equilibrium, or projection along the line of sight \citep[][]{Mark06}. 
Corresponding pre-shock pressure distribution has a broad peak near  $1.6\times 10^{-3}$ keV cm$^{-3}$.

Joint model-constraints on the Mach number are shown in the Fig. \ref{fig:final} (top panel). 
While the SZ measurement supports a lower range of Mach numbers with peak likelihood below $\mach < 2$, the X-ray prior sets it to a higher value due to the low ambient pressure. To put this low pressure region in perspective, the equivalent cluster mass at the shock center for producing a similar pre-shock pressure from the \citet{Say13} GNFW model will be $M_{500,c} \approx 4\times 10^{14}$ M$_{\odot}$, which is roughly 4.5 times lower than the X-ray or weak-lensing based mass estimates for El Gordo. 
Combined results thus point toward a stronger shock, with $\mach = 2.4^{+1.3}_{-0.6}$ (Fig. \ref{fig:final}, top panel inset), 
very similar to the radio measurement of \citetalias{Lind14} based on the 
theory of diffusive shock acceleration (\citealt{Bland87}, \citealt{Kang12}).

\subsection{Radio modeling and the relic $B$-field}
\label{sec:radmodel}

From the Mach number and the upstream gas temperature one can estimate the relic magnetic field, assuming the relic's width to be roughly equal to the product of the velocity and the radiative lifetime of the non-thermal electrons 
(see \citealt{vanWe10}, \citetalias{Lind14}). 
The high-resolution radio image enables us to model a deprojected synchrotron emissivity profile, shown in the 
 Fig. \ref{fig:final} (bottom panel). The fit suggests a narrow lognormal shape, with best-fit parameters $[\mu,\sigma] = [2.1,0.9]$ kpc (see \citealt{szrelic}, Equation (9)), and a shock radius  882 kpc that is consistent with the SZ measurement. 
 The observed broadening (Fig. \ref{fig:prof}, top panel) comes from the projection of a 560 kpc diameter relic surface. 
 The three-dimensional width of the relic is $\sim 15$ kpc (at half peak-maximum), to maintain which one would require a high value of the field strength, of the order of $4-10~\mu$G (Fig. \ref{fig:final}, bottom panel inset).  

Such strong fields at such low density are extremely difficult to justify if they result from the amplification of primordial seed fields. Indeed, even in the presence of a well-developed dynamo, the typical field strength in this environment is expected to be at most a few $\mu$G (e.g., \citealt{Vaz15b}, Fig. 2). 
The restricted time available for dynamo amplification at this high redshift exacerbates the problem. 
Alternatively, part of the observed width might be caused by turbulent re-acceleration of the electrons in the downstream region (\citealt{Fuj15}) that coincides with the location of El Gordo's radio halo  
or by small-scale dynamo amplification of the downstream magnetic fields \citep[][]{Don16}.


\section{Conclusions}
\label{sec:dis}

We present one of 
the first ALMA measurements of the SZ effect, detecting a shock feature in the famous El Gordo cluster at $z=0.87$. The shock is coincident with the location of a prominent radio relic and count as the highest redshift confirmed detection of a merger shock. Besides the ALMA-SZ data, we also analyze archival \chan X-ray data and an ATCA 2.1 GHz radio image to present a self-consistent picture of the thermal and non-thermal signal variation across the shock front. 
SZ and X-ray data are modeled using a Bayesian technique that illustrates respective parameter degeneracies and show how the ALMA measurement of the shock Mach number can improve from an overall normalization of the SZ signal. A future paper will present new ALMA Compact Array data for this purpose along with more details on our multi-wavelength analysis.

ALMA data alone provide evidence for a shock at more than 98\% confidence from the measurements of an underlying pressure discontinuity. However, the Mach number can be constrained only within a wide range, $\mach = 1.4^{+1.2}_{-0.2}$. A joint SZ/X-ray modeling gives preference for a stronger shock, with $\mach = 2.4^{+1.3}_{-0.6}$. 
Fitting the observed relic width with a synchrotron emissivity model indicates a magnetic field strength of the order of $4-10~\mu$G inside the relic, which is unexpectedly high for this redshift.  

These results make use of a relatively short ALMA measurement (3 hr on-source), demonstrating the tremendous potential of future ALMA-SZ observations to model cluster astrophysics from shocks and other substructures.


\vspace{-2mm}
\acknowledgments
We acknowledge R. Lindner and A. Baker for providing the ATCA radio image, F. Pacaud for an X-ray  analysis, T. Erben for an HST color image, and C. Porciani and R. Sunyaev for discussions.  
We thank the anonymous referee for a positive and helpful report. 
Financial support from the Deutsche Forschungsgemeinschaft (DFG) was received through the following programs and grants: MS, FB (TRR33); JE, FB (SFB956); BM (SPP1573) and FV (VA 876/3-1). 
This paper makes use of the following ALMA data:
ADS/JAO.ALMA\#2015.1.01187.S . ALMA is a partnership of ESO (representing
its member states), NSF (USA) and NINS (Japan), together with NRC
(Canada), NSC and ASIAA (Taiwan), and KASI (Republic of Korea), in
cooperation with the Republic of Chile. The Joint ALMA Observatory is
operated by ESO, AUI/NRAO and NAOJ. 
This ALMA project was supported by the German ARC node.


\vspace{2mm}
\facilities{ALMA, CXO, ATCA}


\end{document}